\documentclass[aps,amsmath,amssymb,twocolumn,preprintnumbers,nofootinbib]{revtex4-1}
\usepackage{amsmath,amssymb}
\usepackage{graphicx}
\usepackage{multirow}
   \usepackage{physics} 
\usepackage{acronym}
\usepackage{hyperref}

\newcommand{\beq}{\begin{eqnarray}}
\newcommand{\eeq}{\end{eqnarray}}

\begin{document}

\title{ Interplay of social distancing and border restrictions for pandemics (COVID-19) \\~\\ via \\~\\the epidemic Renormalisation Group framework}


\author{Giacomo Cacciapaglia}
\email{g.cacciapaglia@ipnl.in2p3.fr}
\affiliation{Institut de Physique des 2 Infinis (IP2I),
CNRS/IN2P3, UMR5822, 69622 Villeurbanne, France}
\affiliation{Universit\' e de Lyon, Universit\' e Claude Bernard Lyon 1, 69001 Lyon, France.}
   \author{Francesco Sannino}
\email{sannino@cp3.sdu.dk}
\affiliation{CP3-Origins \& the Danish Institute for Advanced Study. University of Southern Denmark. Campusvej 55, DK-5230 Odense, Denmark; \\~\\
\mbox{ Dipartimento di Fisica E. Pancini, Universit\`a  di Napoli Federico II | INFN sezione di Napoli}\\ \mbox{Complesso Universitario di Monte S. Angelo Edificio 6, via Cintia, 80126 Napoli, Italy.}}

\begin{abstract}
We demonstrate that the epidemic renormalisation group approach to pandemics  provides an effective and simple way to investigate the dynamics of disease transmission and spreading across different regions of the world. The framework also allows for reliable projections on the impact of travel limitations and social distancing measures on global epidemic spread. We test and calibrate it on reported cases while unveiling the mechanism that governs the delay in the relative peaks of newly infected cases among different regions of the globe.  We discover that social distancing measures are more effective than travel limitations across borders in delaying the epidemic peak. 
We further provide the link to compartmental models such as the simplistic and time-honoured SIR-like models. We also show how to generalise the framework to account for the interactions across several regions of the world, replacing or complementing large scale simulations.
 \end{abstract}

\maketitle

\section{Introducing the framework}
The COVID-19 pandemic is raging around the world with an immense toll in terms of human, economic and social impact. Forecasting a pandemic dynamics and its spreading is therefore paramount in helping governments to make informed decisions on a number of social and economic measures, apt at curbing the pandemic and dealing with its aftermath. 

While different empirical models already exist to describe the epidemic dynamics locally and globally, a coherent framework is missing. Using a powerful language and methodology borrowed from high energy physics, we study and forecast  the spreading dynamics and containment across different regions of the world.  This framework is the renormalisation group approach \cite{Wilson:1971bg,Wilson:1971dh}, which was successfully employed for epidemic dynamics in  \cite{Sannino:2020epi}. Here we will generalise the framework to take into account the dynamics in between different regions of the world. The approach is complementary to other  methods  summarised in \cite{LI2019566,ZHAN2018437,Perc_2017, WANG20151,WANG20161}. As for the  widely adopted choice to represent the data by fitting them to simple minded logistic functions we refer to \cite{Danby85,Brauer2019,Miller2012,Murray,Fisman2014,Pell2018}.   We will also provide a map between our framework and compartmental models such as  the time-honoured SIR model \cite{Kermack:1927}.  Our \emph{epidemic renormalisation group (eRG)} approach is based upon a  simpler set of equations, which can be extended in a straightforward way to include interactions between multiple regions of the world, without the need for powerful numerical simulations.

As already noted in \cite{Sannino:2020epi}, rather than the number of cases, it is convenient to discuss its logarithm, which is a more slowly varying function. We define through it an epidemic strength function
\begin{equation} 
\alpha(t) = \rm ln\ \mathcal{I}(t) \ ,
\end{equation} where $\mathcal{I} (t)$ is the total number of infected cases {\it per million} inhabitants in the region, and $\ln$ indicates its natural logarithm.  The derivative of $\alpha$ with respect to time provides a new quantity that we interpret as the {\it beta-function} of an underlying microscopic model. In statistical and high energy physics, the latter governs the time (inverse energy) dependence of  the interaction strength among fundamental particles. Here it regulates infectious interactions.

 More specifically, as the renormalisation group equations in high energy physics are expressed in terms of derivatives with respect to the energy $\mu$, it is natural to identify the time as
  $t/t_0=-\ln {\mu/\mu_0}$, where $t_0$ and $\mu_0$ are respectively a reference time and energy scale. We choose $t_0$ to be one week so that time is measured in weeks, and will drop it in the following. 
 Thus, the dictionary between the eRG equation for the epidemic strength $\alpha$ and the high-energy physics analog is
 \begin{equation}
 \beta (\alpha) = \frac{d \alpha}{d\ln\left(\mu/\mu_0\right)} = - \frac{d \alpha}{dt} \ . 
\end{equation} 
 It has been shown in \cite{Sannino:2020epi} that $\alpha$ captures the essential information about the infected population within a sufficiently isolated region of the world.
The pandemic beta function can be parametrised as
\beq
- \beta (\alpha) = \frac{d \alpha}{dt} = \gamma \, \alpha   \left( 1 - \frac{\alpha}{a} \right)^n\,,
\label{eq:beta0}
\eeq 
 whose solution, for $n=1$, is a familiar logistic-like function
\beq
 \alpha (t) =  \frac{a e^{\gamma t}}{b + e^{\gamma t}}\,.
\eeq
The dynamics encoded in Eq.~\eqref{eq:beta0} is that of a system that flows from an UV fixed point at $t=-\infty$ where $\alpha = 0$  to an IR fixed point where $\alpha = a$. The latter value encodes the total number of infected cases per million expected in the region under study. The coefficient $\gamma$ is the diffusion slope, while $b$ shifts the entire epidemic curve by a given amount of time. Further details, including what parameter influences the {\it flattening of the curve} and location of the inflection point and its properties can be found in \cite{Sannino:2020epi}. Note that here we work with number of cases per million, so that our $\alpha$ corresponds to $\alpha - \ln n_m$  of  \cite{Sannino:2020epi} with $n_m$ the number of inhabitants per million per each sufficiently isolated region of the world. 
 In this work, we extend the eRG formalism to include the diffusion of the epidemic between multiple nearly-isolated regions. 

Our work is organised as follows.
In Sec.~\ref{sec:2} we present the formalism in the simple case of two regions, and study how the interaction term influences
the delay between the epidemic peak between the first and second regions. We also investigate the effect of closing the borders
at different times between the two regions.
In Sec.~\ref{sec:3} we map the eRG formalism onto time-honoured compartmental models  of the SIR-type. 
In Sec.~\ref{sec:4} we test the eRG formalism by comparing the predictions to data relative to the COVID-19 epidemic in Europe and
in the United States, and generalise it to include multiple regions at the same time. We then use the model to simulate a second wave of
epidemic diffusion in a sample of European countries.
Finally, we present a discussion of the results in Sec.~\ref{sec:5} and offer our conclusions.


\section{Epidemic diffusion among different regions of the globle}
\label{sec:2}

Here we go beyond the state-of-the-art by considering the diffusion among multiple regions of the world, each characterised by its own $\alpha_i (t)$, which in isolation
obeys a beta function like Eq.~\eqref{eq:beta0}, with its own $\gamma_i$ and $a_i$.

We exemplify the framework by first considering two regions, and we generalise to multiple ones later. 
To couple the two equations, we start from the following axiom: there is a constant number of travellers moving from one region to 
the other, and viceversa, given by $\Delta N_{\rm trav}$ each week.
Our basic simplifying assumption is that the number of travellers is symmetric, i.e. there is no net flow of people between the two regions:
this is a reasonable approximation during a short time as immigration only involves a smaller fraction of inhabitants than that involved in 
the epidemic.
We further use the approximation that the rate of infected cases within the travelling subset of people is the same as the rate
of infected cases in the total population of each region. Thus, the variation in the number of infected cases per million in region-1 
is given by~\footnote{Here we neglect the fact that part of the infected population has recovered, thus probably ceasing to be infectious.
We will come back to discussing the validity of this approximation.}
\beq
n_{m1} \frac{\delta \mathcal{I}_1 (t)}{\delta t} = k \left( \mathcal{I}_2 (t) - \mathcal{I}_1 (t) \right)\,,
\label{eq:deltaI}
\eeq
where $n_{m1}$ is the population of region-1 in millions and
\beq
k = 10^{-6}\, \Delta N_{\rm trav}\,.
\eeq
For region-2, we find the analogous
\beq
n_{m2} \frac{\delta \mathcal{I}_2 (t)}{\delta t} = k \left( \mathcal{I}_1 (t) - \mathcal{I}_2 (t) \right) \,,
\label{eq:deltaI2}
\eeq
where the same $k$ applies. Physically, the parameter $k$ measures the number of reciprocal travellers per week in units of million people.
For instance, if the number of weekly travellers is $\Delta N_{\rm trav} = 1000$, then $k = 10^{-3}$.
Using the identity
\beq
\frac{\delta \mathcal{I}_i (t)}{\delta t} = \mathcal{I}_i \frac{\delta \alpha_i}{\delta t}\,,
\eeq
the effect of this exchange can be encoded in the two beta functions, c.f. Eq.~\eqref{eq:beta0}, as follows:
\beq
- \beta (\alpha_1) &=& \gamma_1 \alpha_1 \left( 1 - \frac{\alpha_1}{a_1} \right) + \frac{k}{n_{m1}} \left( e^{\alpha_2 - \alpha_1} -1 \right)\,, \label{eq:beta1} \\
- \beta (\alpha_2) &=& \gamma_2 \alpha_2 \left( 1 - \frac{\alpha_2}{a_2} \right) + \frac{k}{n_{m2}} \left( e^{\alpha_1 - \alpha_2} -1 \right)\,. \label{eq:beta2}
\eeq

The above equations   describe the evolution of the epidemic across the two regions, once a small fraction of the population
travels between the two. However, for large $k$, they have the interesting property of forcing $\alpha_1 = \alpha_2$, which in turn modifies 
the value of the fixed point for the two regions. The fact that the $\alpha$'s become equal in the long run indicates that the two regions have merged
into one. One surprising finding is that the total number of infected cases across the two regions, for large $k$, may be reduced compared to the isolated case
(see details in Appendix~\ref{app:largeK}).
While mathematically intriguing, we do not consider this result physical, as having large $k$ modifies the values of $\alpha_i$ and $\gamma_i$
in the two regions compared to the values one would have in case of isolation. In other words it would violate our initial assumption that the two regions are nearly-isolated,
with small $k$. One can go beyond the realistic case envisioned here by increasing $k$. This would require modifying the set of equations substantially and goes beyond the scope of this work.  

\begin{table}[ht]
\begin{center}
\begin{tabular}{l|cc|c|}
Region   & \multicolumn{2}{c|}{Per million} &  \\
             &  $a$  &  $\gamma$  & $n_m$\\
\hline
China (Hubei)  &  $7.22$ & $0.97$   & $59$ \\
South Korea  &$5.25$ & $1.29$ &  $51$\\
United States & $8.41$ & $0.47$  & $331$\\
France &$7.92$   & $0.69$   & $65$ \\
Spain & $8.61$ &  $0.57$ &  $47$\\
Italy  & $8.22$  &  $0.49$  &  $60$\\
United Kingdom & $8.17$ &  $0.48$ & $68$\\
Germany & $7.61$ &  $0.68$ &   $84$\\
Switzerland & $8.15$ & $0.76$  & $8.7$\\
Denmark & $7.55$ &  $0.47$ &  $5.8$\\
\hline
 \end{tabular} 
\caption{Fits for individual countries, assuming they are isolated systems \cite{Sannino:2020epi}.
We use data updated to the 4th of May (from {\tt www.worldometers.info}), so the values should be considered
as averaged over the whole period of pandemic diffusion.}
\label{tab:agamma}
 \end{center} 
\end{table}
To quantitatively estimate the interaction between two regions of the world, we consider benchmark values for the parameters in the two beta functions using the results given in \cite{Sannino:2020epi}.  We show in Table~\ref{tab:agamma} the values of $a$ and $\gamma$ for various regions of the world
for the COVID-19 pandemic \cite{Sannino:2020epi}. $a$ is normalised per million inhabitants and all values are adjourned to the 4th of May, 2020.~\footnote{It is 
straightforward to provide daily updates, as done following the eRG approach \cite{Sannino:2020epi} at {\tt http://caracal.imada.sdu.dk/corona/} for different regions of the world.}
The values of $\gamma$ and $a$ are average values over the whole duration of the epidemic diffusion in each
country/region. We observe that the value of $\gamma$ tends to diminish over time as a consequence of the effect of gradual implementing of social distancing
measures in each region. At the early stages of the epidemic, we observe $\gamma \sim 1$, so that we will consider this as a benchmark value for
the epidemic diffusion without any restriction.  

With the exception of South Korea and China (Hubei province), the range for $a$ is roughly $[7.5, 8.6]$, while for $\gamma$ we find $[0.4,0.76]$. Thus, we defined the following benchmark scenario for the two regions:
\beq
 \begin{array}{c}  a_1 = 7\,, \;\; b_1 = 2\,, \\
  a_2 = 8\,, \;\; b_2 = 200,\, \infty \,; \end{array}
 \label{eq:benchmark0} 
\eeq
while we vary the values of $\gamma_1$ and $\gamma_2$ as specified in the figures. 
The value of $b_2=200$ is chosen such that the peak in the two regions in isolation have a relative delay of 14 weeks. 
The peak is here defined as the week where the maximum number of new infected cases per million is registered and corresponds to the inflection point of the total number of infected cases curve. The explicit formula for the inflection point time as function of the parameters of the theory can be found in \cite{Sannino:2020epi}. 

\begin{figure}[tb]
\begin{center}
\includegraphics[width=7cm]{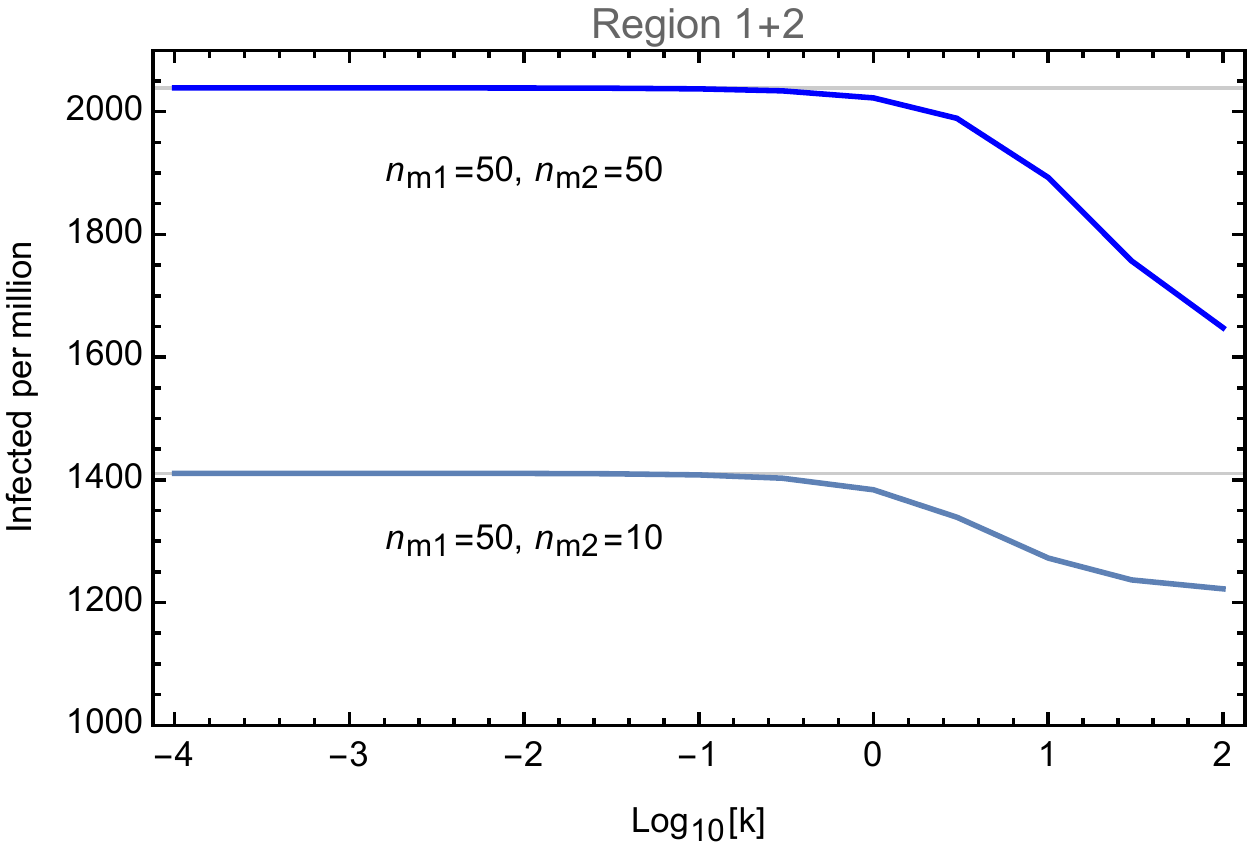}
\end{center}
\caption{Infected cases per million in the sum of the two regions, as a function of $k$ and for two choices of the region population.} \label{fig:total}
\end{figure}

As a first sanity check, we computed the total number of infected cases across the two regions, per million, at the end of
the pandemic, i.e. at infinite time. This is given by
\beq
\mathcal{I}_{1+2} (\infty) = \frac{n_{m1} \mathcal{I}_1 (\infty) + n_{m2} \mathcal{I}_2 (\infty)}{n_{m1} + n_{m2}}\,,
\eeq
as a function of $k$. The result allows us to determine the largest value of $k$ that does not affect the total number, i.e.
the largest value that $k$ can have before the two regions effectively merge into one. In Fig.~\ref{fig:total} we show the results for two different populations. The plot shows that $k$ as large as $0.1$ is allowed before our description of the coupled system breaks down. 
Note that the maximal value of $k$ grows linearly with the population in region-2, as it enters as the ratio $k/n_{m2}$ in the coupled differential equations.

\subsection{Peak delay study}

To understand how the interaction encoded by $k$ affects the diffusion of the epidemic in the two regions, we study the same
benchmark of Eq.~\eqref{eq:benchmark0}, except that we set $b_2 = \infty$, i.e. the region-2 remains with zero infected cases
if isolated.  
One caveat that should be kept in mind is that the values for $\gamma$ in Table~\ref{tab:agamma} are obtained by fitting the data
during the whole period of the epidemic, i.e. they take into account the effect of social distancing measures in each region.
However, at the early stages of the epidemic, when social distancing measures were not yet being enforced, we expect larger values of $\gamma$. That is the reason why, in the following, we assume $\gamma = 1$
as the natural initial benchmark value. Nevertheless, we  show how different social distancing measures impact the results for region-2 by showing also the results for smaller values of $\gamma_2$.

\begin{figure}[tb]
\begin{center}
\includegraphics[width=7cm]{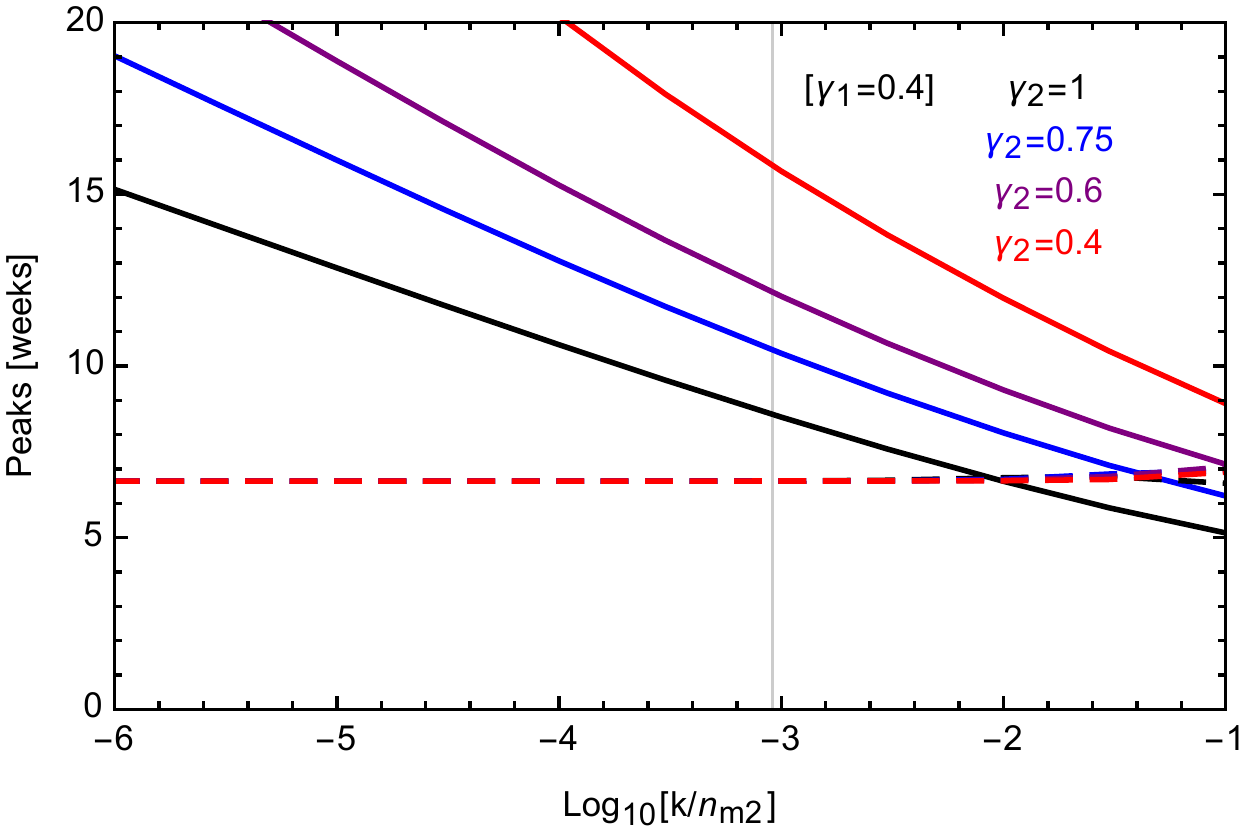} 
\vspace{0.5cm}
\includegraphics[width=7cm]{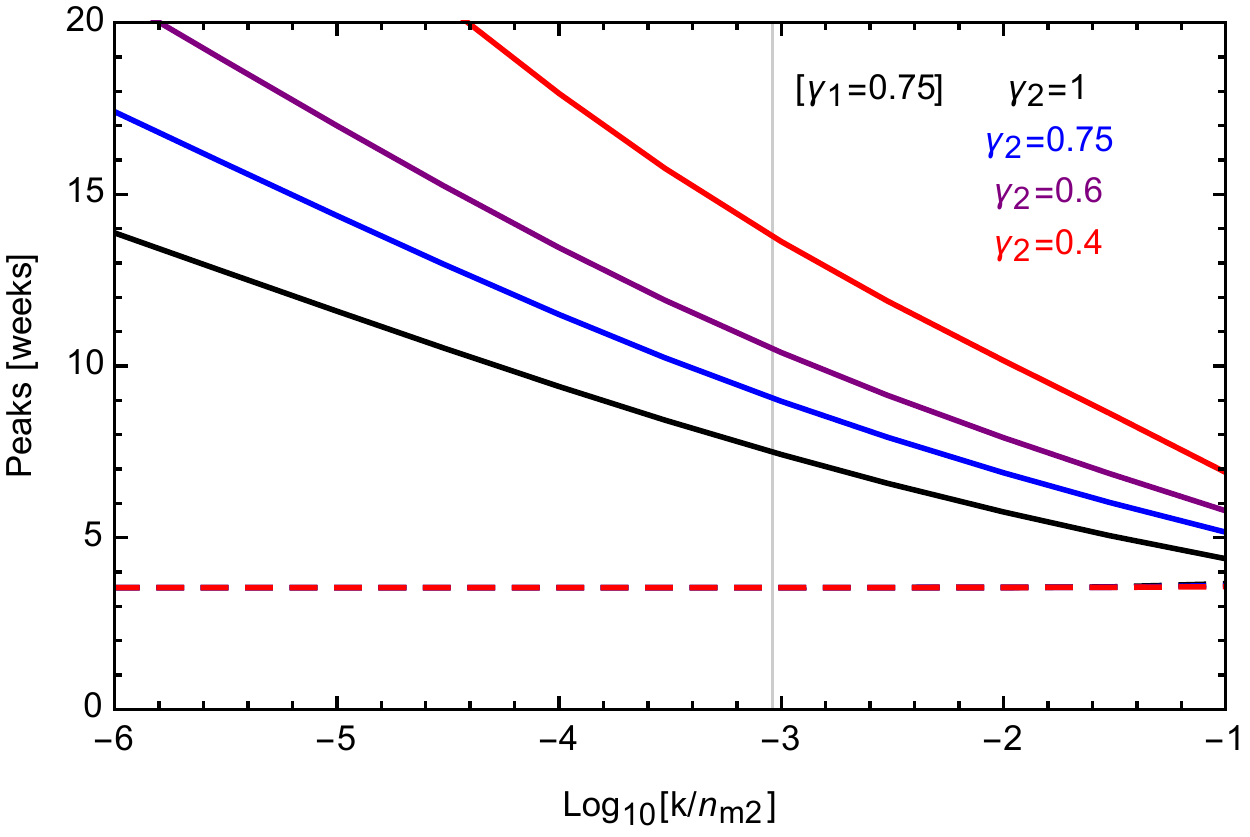}
\end{center}
\caption{Peak timing for region-1 (dashed) and region-2 (solid) for $\gamma_1 = 0.4$ (top) and $\gamma_1 = 0.75$ (bottom), and different values for $\gamma_2$.} \label{fig:delay}
\end{figure}

We discover that the interaction among the two regions of the world,  controlled by the parameter $k$, is sufficient to ignite the spread of the epidemic to region-2 and it also controls the timing of the peak. This is shown in Fig.~\ref{fig:delay}, where we plot the time of the peaks in the two regions
as a function of $k/n_{m2}$. The result does not depend on $n_{m1}$. Also, the time of the peak
for region-1 is unaffected by the value of $k$ (dashed curve)  while it affects the timing of the peak for region-2 (solid curves). 
Note that the $k$-term in Eq.~\eqref{eq:beta2} sparks the epidemic diffusion in region-2 as soon as
$\frac{k}{n_{m2}} e^{\alpha_1 (t)}$ becomes sizeable. After this point, the epidemic evolution follows the solution of the initial equation
\eqref{eq:beta0}, as encoded in the first term of the region-2 beta function. 

The numerical results for the peak delay show a linear dependence on $\ln k$, with a change in slope appearing for $k/n_{m2} \sim 10^{-3}$.
This value corresponds to
\beq
\frac{k}{n_{m2}} e^{a_1} = 1\,,
\eeq
i.e. it marks the threshold (grey line in the plots) between the regime where the interaction term is always smaller than one, and the one where strength 1
is attainable. 

To test our approach we consider the COVID-19 epidemic spread from China (Hubei province)
to Europe (Italy). From data it is  known that the peaks in the two regions are about 7 weeks apart. A reasonable estimate of weekly travellers between the two regions
is in the order of the thousands, so we consider $k = 5 \times 10^{-3}$ as a benchmark. This means that
\beq
\frac{k}{n_{m2}} \sim 10^{-4}\,.
\eeq
For this value, the bottom plot in Fig.~\ref{fig:delay} allows us to estimate the peak delay to be around $6$ weeks for $\gamma_2 = 1$, i.e. for unrestricted diffusion within Italy. This nicely confirms our expectations while validating the model. 

It is useful to note that both $k$ and $b_2$  lead to a temporal shift of the epidemic curve for region-2, however the underlying mechanisms are distinct. The former is due to an interaction between two different regions of the world, while the latter is a constant of integration that depends on the number of cases at the initial time $t=0$ in region-2. This also means that a specific peak time for region-2 relative to region-1 can emerge as a combination of the two effects, interpolating between the two limiting cases: the peak delay is entirely due to the interaction with region-1, or it is due to the presence of cases in region-2 (which may have different origin) and the coupling to region-1 is negligible. We will discuss this interplay in more details in the next section.

\subsection{Border control versus social distancing}

\begin{figure}[tb]
\begin{center}
\includegraphics[width=7cm]{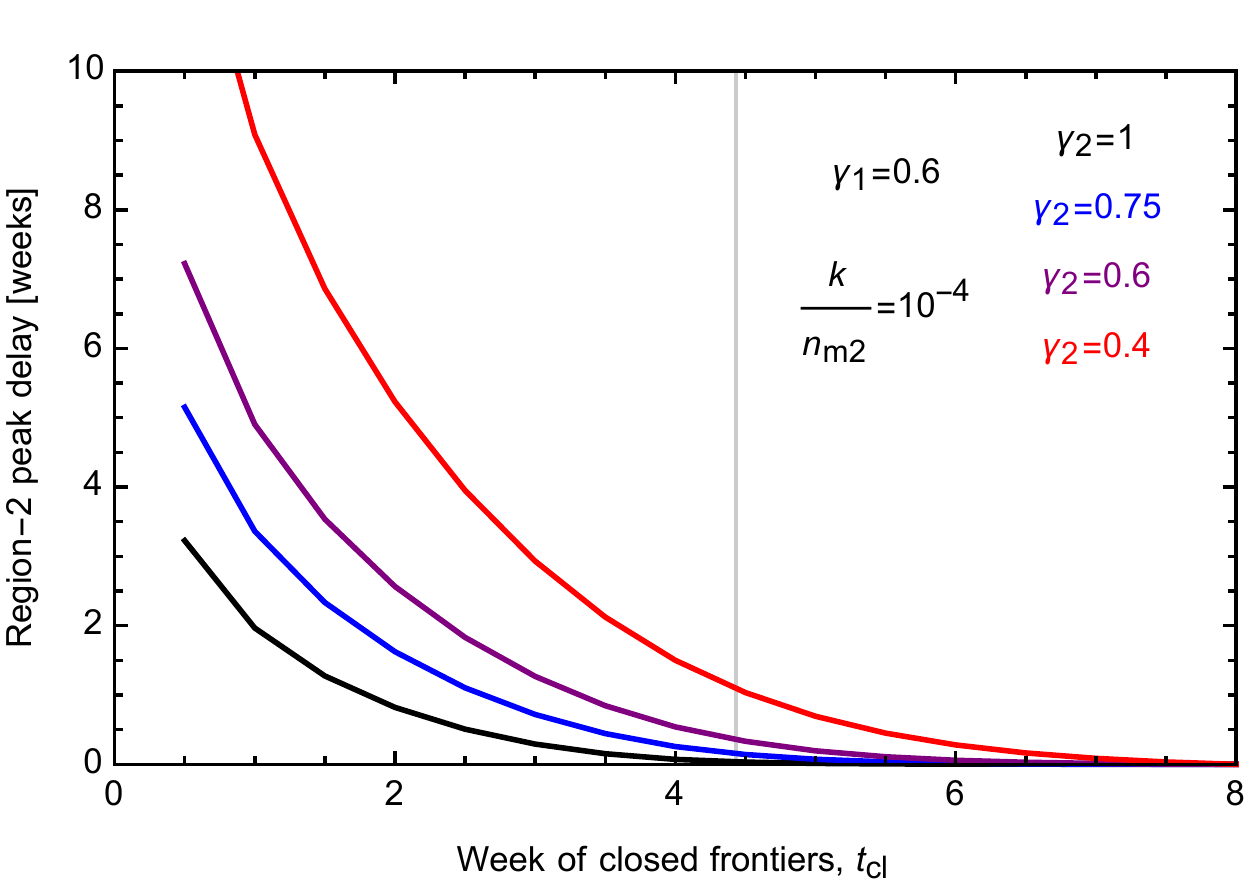} \hspace{0.7cm}
\includegraphics[width=7cm]{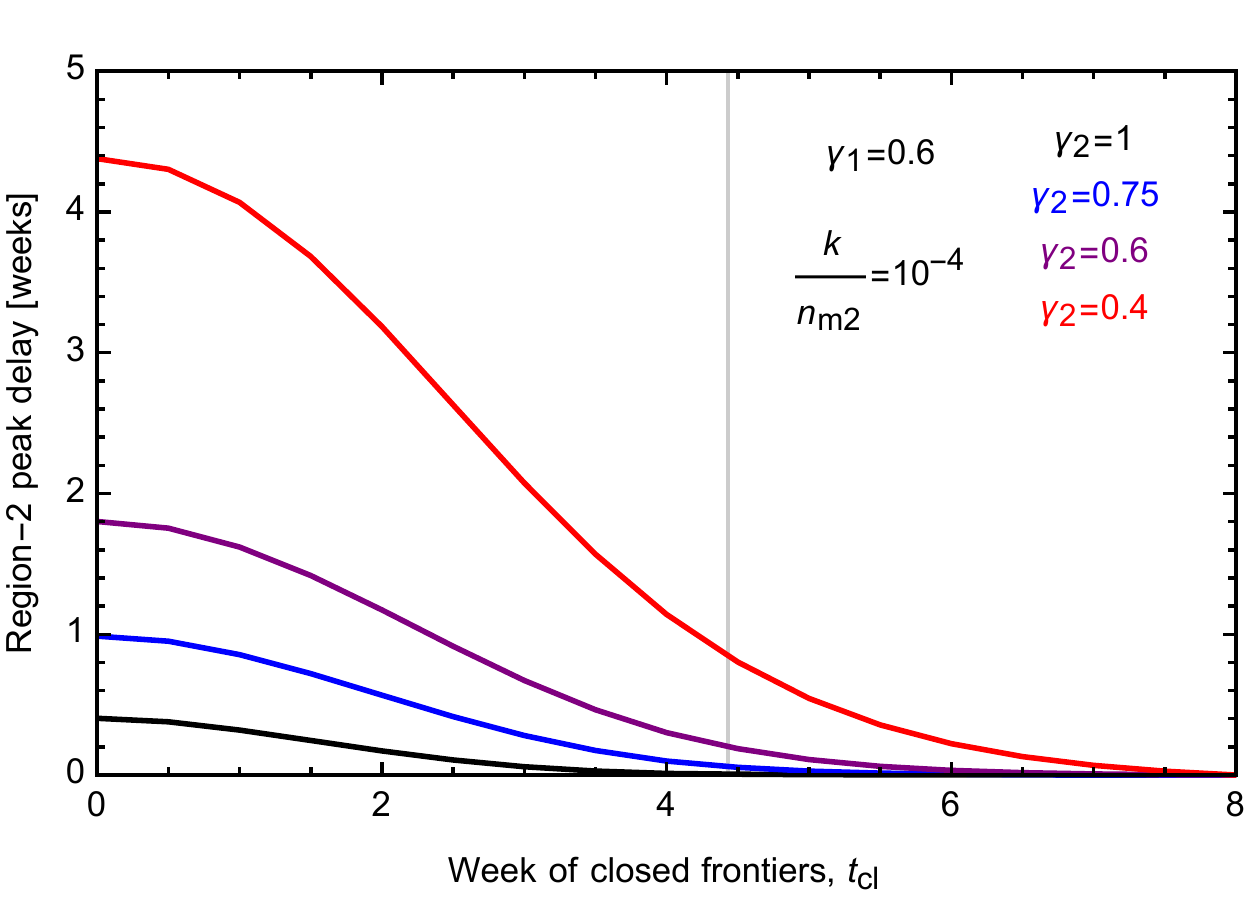}
\end{center}
\caption{Delay in the epidemic peak in region-2 for fixed $k/n_{m2} = 10^{-4}$, as a function of the week $t_{\rm cl}$ when the borders are closed (after which $k=0$). The top panel corresponds to $b_2 = \infty$ (no initial cases), while the bottom one to $b_2 = 200$. The vertical line marks the time of the peak for region-1.} \label{fig:delay2}
\end{figure}

We now turn out attention on the impact of closing the borders between two regions of the globe versus different degree of social distancing. In the eRG approach this is  implemented by setting to zero $k$ after the closing time $t_{\rm cl}$.
We consider the benchmark values given in Eq.~\eqref{eq:benchmark0}, while the impact of social distancing is encoded in region-2 in varying the value for $\gamma_2$. Furthermore we consider two scenarios: one in which region-2 has zero initial cases, meaning that the epidemic would not occur for $k=0$ (corresponding to $b_2 = \infty$) and another where we fix the initial condition according to the benchmark (corresponding to $b_2 = 200$).

The results are shown in Fig.~\ref{fig:delay2}, where we report the delay in the peak of region-2 caused by closing the borders  (i.e., delay relative to the case of $t_{\rm cl} = \infty$). Such a delay depends crucially on the value of $\gamma_2$  in region-2 as shown in the top plot when the epidemic in region-2 is only driven by the interaction term. In particular the results show that a significant delay in the spreading of the epidemic can be achieved only if the closing is enacted before the peak in region-1 (which is unaffected by $k$). 

In the bottom panel we show the case where region-2 features already some initial cases, so that $t_{\rm cl}=0$ would correspond to isolated regions with both featuring infected cases. In this case, we also see that the effect of the interaction is more pronounced for small values of $\gamma_2$ in region-2, indicated by the red curve for $\gamma_2 = 0.4$. For this value of $\gamma_2$, isolation would yield a delay of $4.5$ weeks in the peak. For larger values of $\gamma_2$ (less social distancing) the  peak delay is strongly reduced to within one or two weeks. In any case, closing the borders is only relevant if done before the peak in region-1 is attained. 

Our results, obtained using the simple and effective eRG approach, agree qualitatively with the ones presented in \cite{Chinazzi} obtained using a numerical analysis. The take home message is  that social distancing plays the dominant role in curbing and delaying the epidemic spread in region-2 with respect to seed region-1.

\section{Relation to the SIR model}
\label{sec:3}

Epidemic dynamics is often described in terms of simplistic compartmental models introduced long time ago in \cite{Kermack:1927}. Here, the affected population is described in terms
of compartmentalised sub-populations that have different roles in the dynamics. Then, differential equations are designed to describe the
time evolution of the various sub-populations. For an application to the COVID-19 epidemic, see~\cite{SEIR,scala2020}. The sub-populations
can be chosen to represent (S)usceptible, (I)nfected and (R)ecovered individuals (SIR model), obeying the following differential equations:
\beq
\frac{d S}{d t} &=& - \tilde{\gamma}\ S \frac{I}{P}\,, \\
\frac{d I}{d t} &=& \tilde{\gamma}\ S \frac{I}{P} - \epsilon\ I\,, \\
\frac{d R}{d t} &=& \epsilon\ I\,;
\eeq
where $P = S + I + R$ is a constant, measuring the total number of individuals affected. As the equations do not depend on the normalisation
of the number of individuals, we can consider them for cases per million. Due to the constant $P$, only two equations are independent, so that we 
can drop the one for $S$. The total number of infected, $\mathcal{I} (t)$, we study in our model is related to the above sub-populations as
\beq
\mathcal{I} (t) = I(t) + R(t)\,.
\eeq
We can therefore re-write the two independent SIR equations as
\beq
\frac{d \mathcal{I} (t)}{d t} &=& \tilde{\gamma} \left(\mathcal{I} (t) - R(t) \right) \left( 1 - \frac{\mathcal{I}(t)}{P} \right)\,, \label{eq:SIR1}\\
\frac{d R (t)}{d t} &=& \epsilon \left( \mathcal{I} (t) - R(t) \right)\,. \label{eq:SIR2}
\eeq
Eq.~\eqref{eq:SIR1} has a form similar to Eq.~\eqref{eq:beta0}, except for the following: it is written in terms of the total number $\mathcal{I} (t)$ instead of its log $\alpha (t)$; it contains a dependence on the number of recovered cases, $R(t)$. 

Thus, our eRG approach would be equivalent to the SIR model if we could drop the $R(t)$ dependence in the differential equation for $\mathcal{I} (t)$. 
It is conceivable that this is the case: in fact, in Eq.~\eqref{eq:SIR1} we can already see that the second factor drives the number of infected cases to
the fixed point $\mathcal{I} (\infty) \to P \equiv e^{a}$, which corresponds to the IR fixed point in the eRG approach. $R(t)$, instead, is zero at early times and
only grows slowly as long as the recovery rate $\epsilon$ is small, thus its effect should remain negligible once the dynamics of $\mathcal{I} (t)$ is driven
towards the fixed point.  As investigated in~\cite{Sannino:2020epi}, the dynamics of $\mathcal{I} (t)$ and $\alpha (t)$ can be described by the same equation,
as they both are driven to flow between the two fixed points.

\begin{figure}[tb]
\begin{center}
\includegraphics[width=7cm]{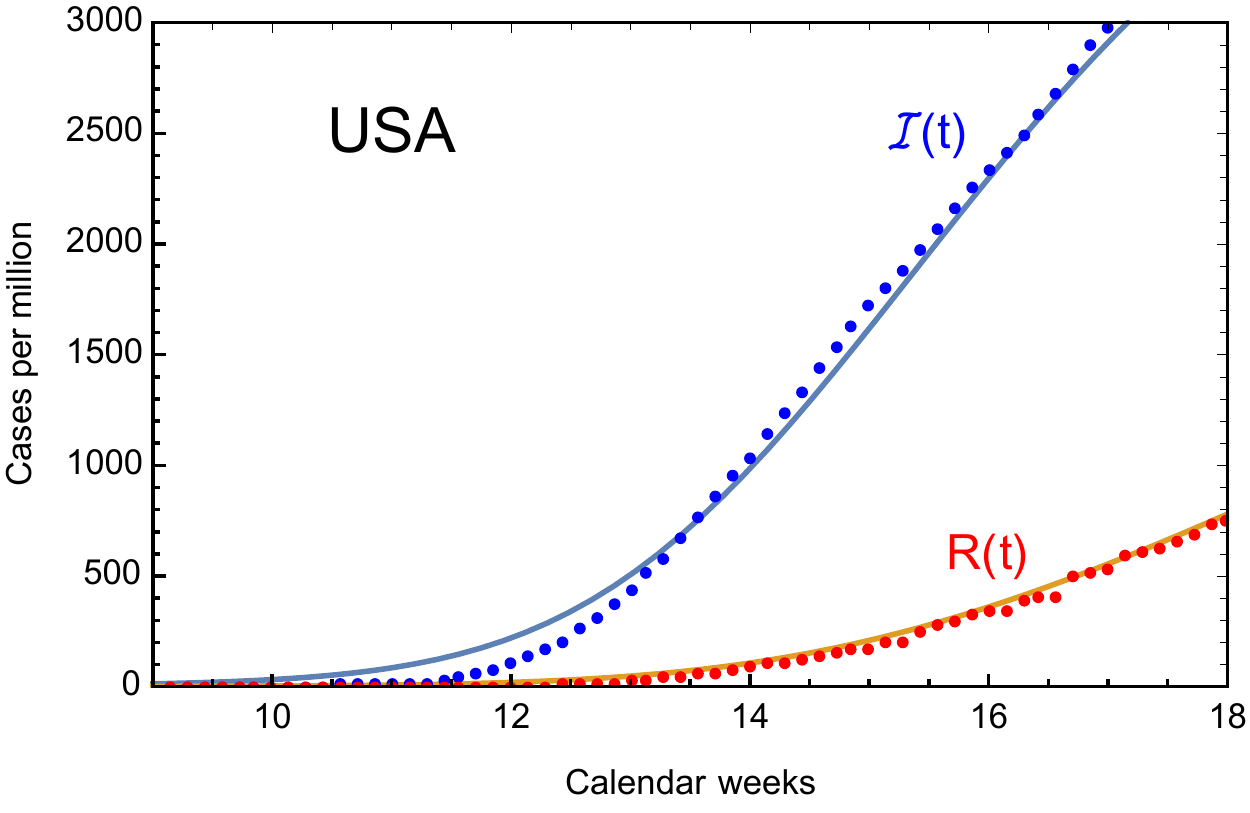}
\end{center}
\caption{Infected $\mathcal{I} (t)$ and recovered $R(t)$ cases per million for the US compared to the data. We used $\epsilon = 0.09$.} \label{fig:USA}
\end{figure}

Once a solution for $\mathcal{I} (t)$ is found following the eRG approach, i.e. Eq.~\eqref{eq:beta0}, the number of recovered cases can be calculated by solving Eq.~\eqref{eq:SIR2}, with solution
\beq
R(t) = \epsilon \int_0^t dx\ e^{\epsilon (x-t)}\ \mathcal{I} (x)\,. 
\label{eq:Roft}
\eeq
To validate this approach and calibrate $\epsilon$, we compared the above formula to the number of recovered cases for the United States (US), where $\mathcal{I} (t)$ is obtained using the fit values in Table~\ref{tab:agamma}: the results are shown in Fig.~\ref{fig:USA}, where $R(t)$ (in red) reproduces the data for $\epsilon = 0.09$. We checked that for other countries, a similarly good fit can be obtained for $\epsilon \sim 0.1$, thus we consider this description consistent.

\begin{figure}[tb]
\begin{center}
\includegraphics[width=7cm]{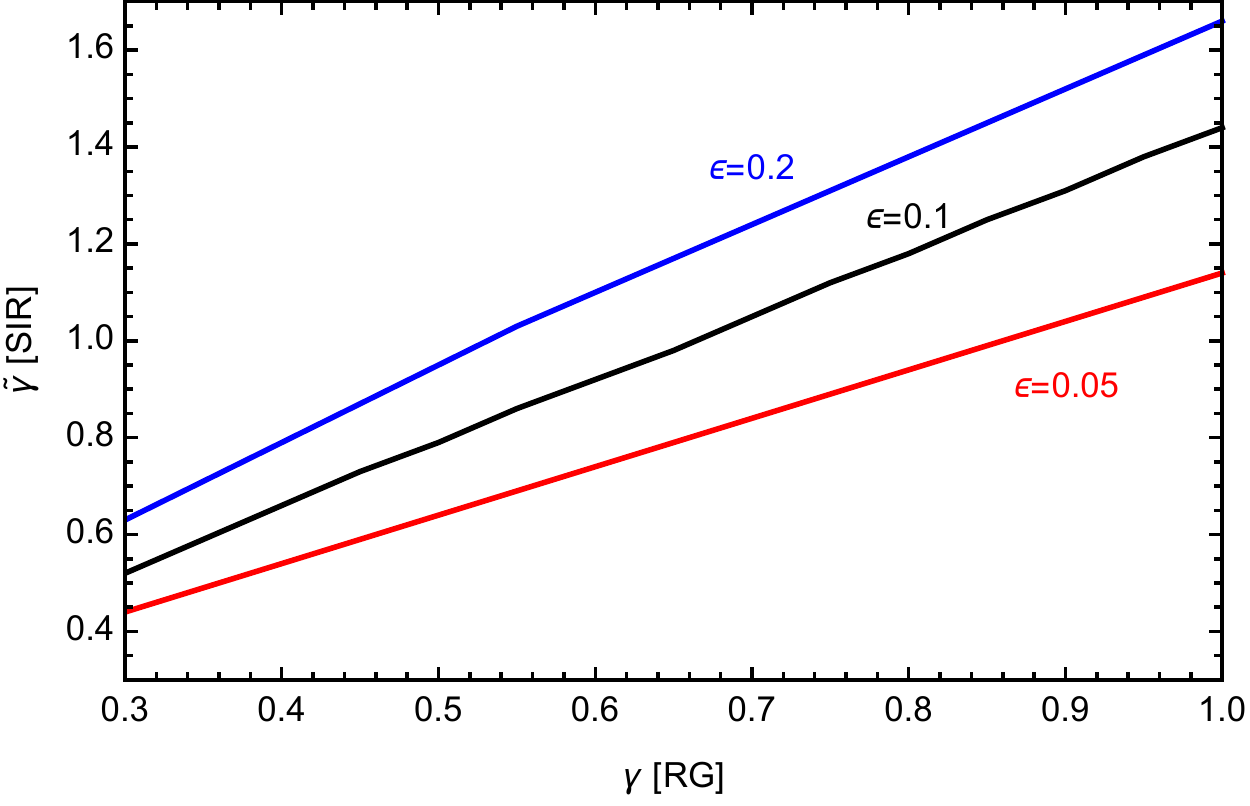}
\end{center}
\caption{Values of $\tilde{\gamma}$ in the SIR model as a function of $\gamma$ in the RG approach, for 3 values of $\epsilon$.} 
\label{fig:gammatilde}
\end{figure}

To establish a more quantitative dictionary between the eRG approach and the SIR model, we compared the numerical solutions of the SIR equations~\eqref{eq:SIR1} and~\eqref{eq:SIR2} to the solutions of the beta function in Eq.~\ref{eq:beta0} (with $R(t)$ given by Eq.~\eqref{eq:Roft}). We find that the solutions overlap as long as 
matching values of $\gamma$ and $\tilde{\gamma}$ are used. In Fig.~\ref{fig:gammatilde} we show the numerical relation between the matching values of the couplings for 3 choices of the recovery rate $\epsilon$: the result shows a linear relation between the couplings in the two models.

Being able to reproduce the number of recovered cases for one region in isolation, we can now address the issue of the effect of the recovered cases
in the coupled system. In fact, the transmission of the epidemic due to travel of individuals between the two regions is only due to the presence of
people actively infected, namely it depends on
\beq
I (t) = \mathcal{I} (t) - R(t)\,.
\eeq
Thus, it suffices to replace the expression in Eq.~\eqref{eq:deltaI} with
\begin{multline}
n_{m1} \frac{\delta \mathcal{I}_1 (t)}{\delta t} = k \left( \mathcal{I}_2 (t) - \mathcal{I}_1 (t)\right) \\
 - k \left(R_2 (t) - R_1 (t) \right) \,,
\end{multline}
and similarly for Eq.~\eqref{eq:deltaI2}. 
We have compared the solutions of the coupled differential equations with and without taking into account the recovered cases, and found that  
including $R_i (t)$ only affects the epidemic diffusion in region-2 by a few days. Thus, this effect can be neglected in first approximation.

\begin{figure*}[tb]
\begin{center}
\includegraphics[width=7cm]{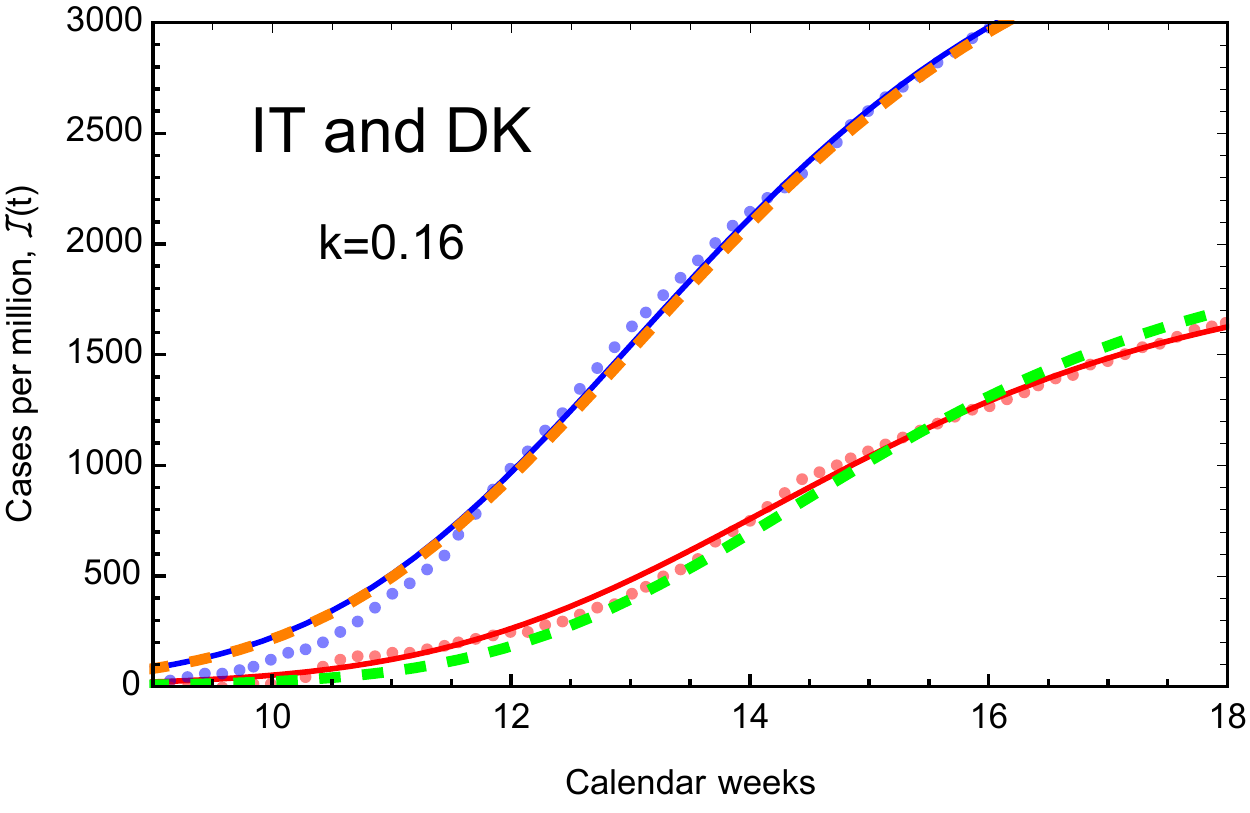} \hspace{1cm}
\includegraphics[width=7cm]{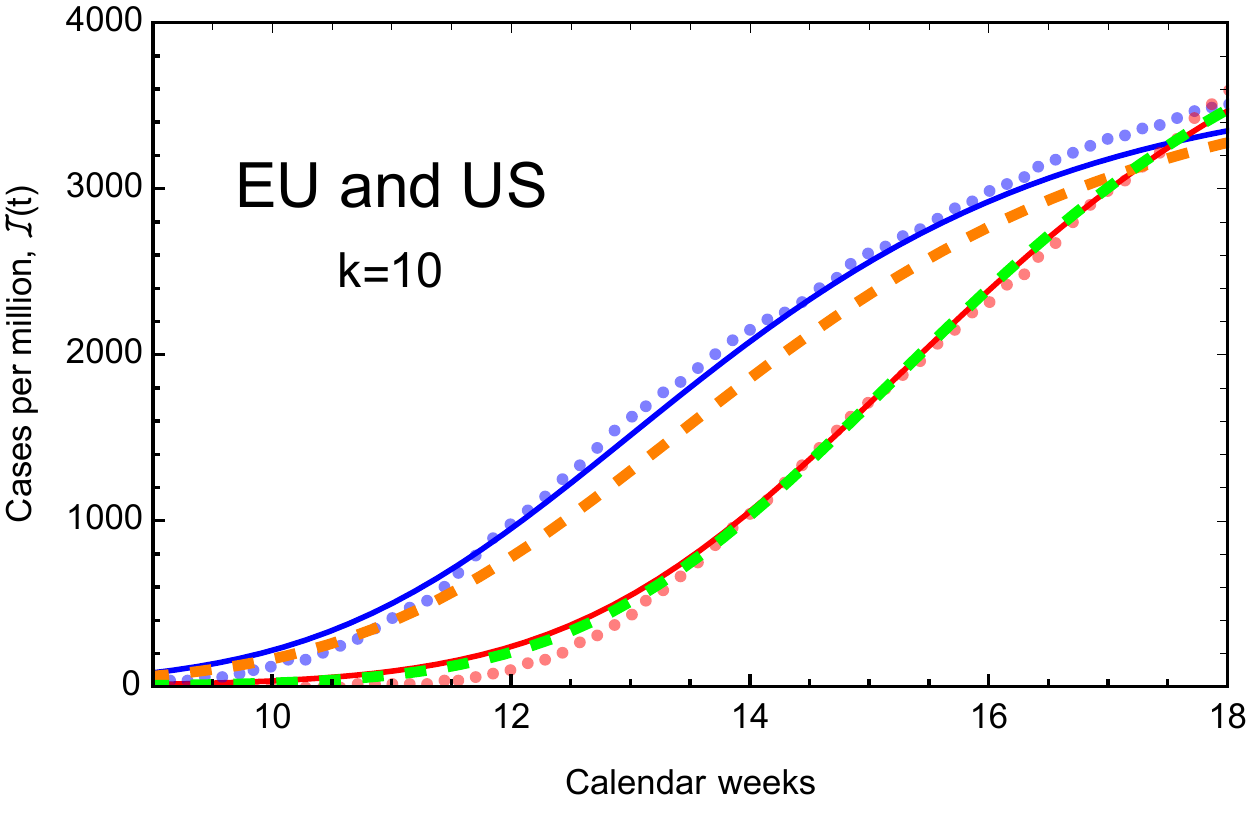}
\end{center}
\caption{Comparison between the infected cases in Italy and Denmark (left plot) and in the European Union and United States (right plot) and the fits in our two-region eRG model.} 
\label{fig:data}
\end{figure*}

\section{COVID-19 examples}
\label{sec:4}

We now  confront the eRG framework to data from the COVID-19 pandemic  collected from {\tt www.worldometers.info} and adjourned to the 4th of May, 2020.  Although we are well aware of the pitfalls stemming from comparing data provided by different countries due to the inhomogeneous way infectious cases were tested and reported, it is still possible to extract from these reliable time behaviour and structure. Of course,  when coupling two regions of the world, part of the initial uncertainty  also affects the epidemic transmission probability without affecting the overall picture. 

Nevertheless, we  will now see that the eRG formalism can be used simultaneously to quantiately project the spreading dynamics across different regions of the world, or as an a-posteriori way to learn how this spreading came to be. We focus on two examples, one intra European  (Italy-Denmark) and the other between Europe and the US. The values for $\gamma$ and $a$ for each country are taken from the fit in Table~\ref{tab:agamma}, which assumes isolation.

\subsection{From Italy to Denmark}

In the right plot of Fig.~\ref{fig:data} we show the total number of infected cases in Italy and Denmark (blue and red dots, respectively), compared to the fit in Table~\ref{tab:agamma} (solid curves): the latter assumes that the epidemic occurred in the two regions while in isolation. We now wish to understand how and whether the virus spread from Italy to Denmark: thus, we used the coupled Eqs~\eqref{eq:beta1} and \eqref{eq:beta2}, while we set the number of initial cases in Denmark to be null, i.e. $b_2 = \infty$. All the other parameters are fixed to the values in the Table. We find that the two curves can be reasonably fit by assuming $k=0.16$, as shown in the left plot of Fig.~\ref{fig:data}: the dashed orange curve, corresponding to Italy, overlaps to the isolated fit (solid blue), while the new curve for Denmark (dashed green) is close to the isolated fit (solid red). Let us now comment on the actual value of $k$.  If we take it literally this it would correspond to a rate of 160.000 travellers between the two regions each week. This is an unreasonably large value but it can be alternatively and  conservatively interpreted in the following ways: 
\begin{itemize}
\item[i)]{More countries contributed to the epidemic spread in Denmark;}
\item[i)]{The original spreading dynamics in Denmark is due to few very socially active infected individuals that traveled back from Italy and/or were super-spreaders;}
\item[iii)] {A combination of the above.}
\end{itemize}
Whatever the reason, it is naturally incorporated in a larger value of $k$. One can also take into account the various scenarios by effectively re-instating an initial value for $\alpha_2$ at $t=0$ while reducing the $k$ value.  

\begin{figure*}[tb]
\begin{center}
\includegraphics[width=7cm]{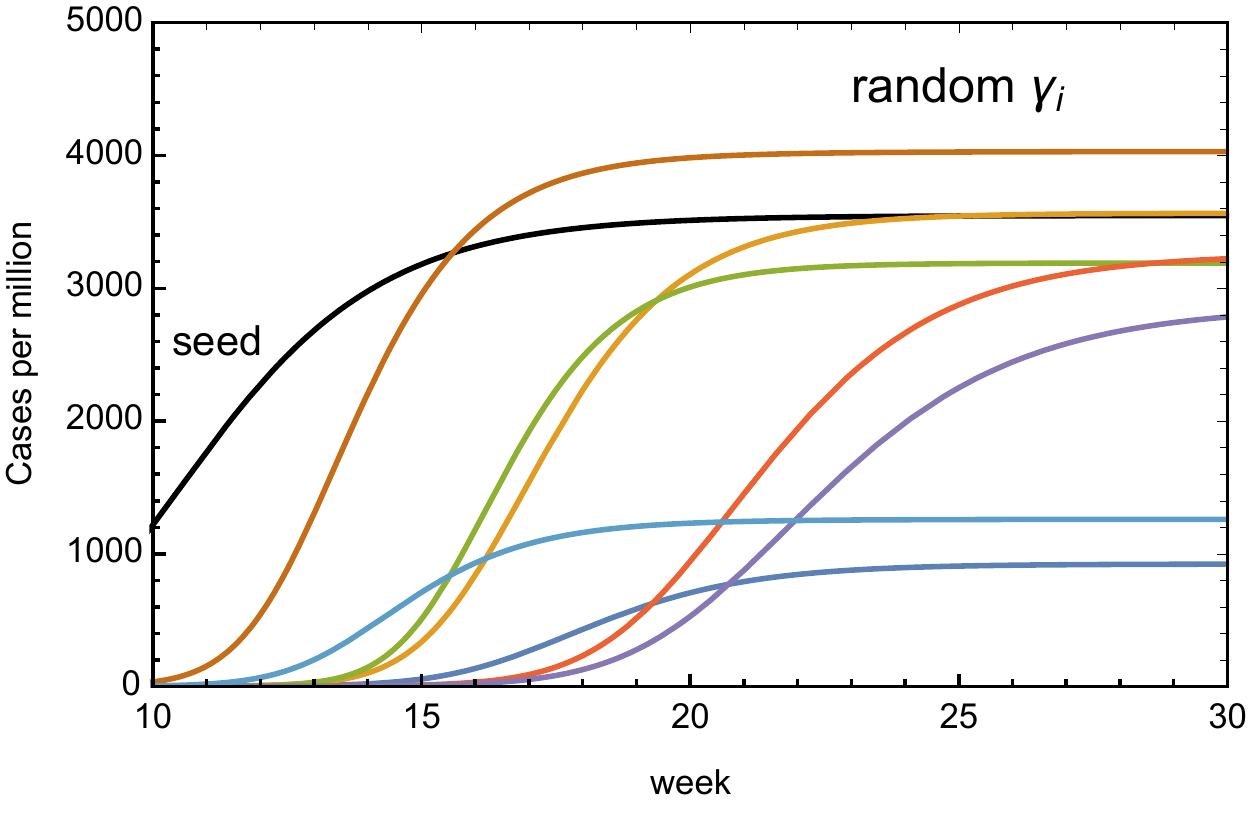} \hspace{1cm}
\includegraphics[width=7cm]{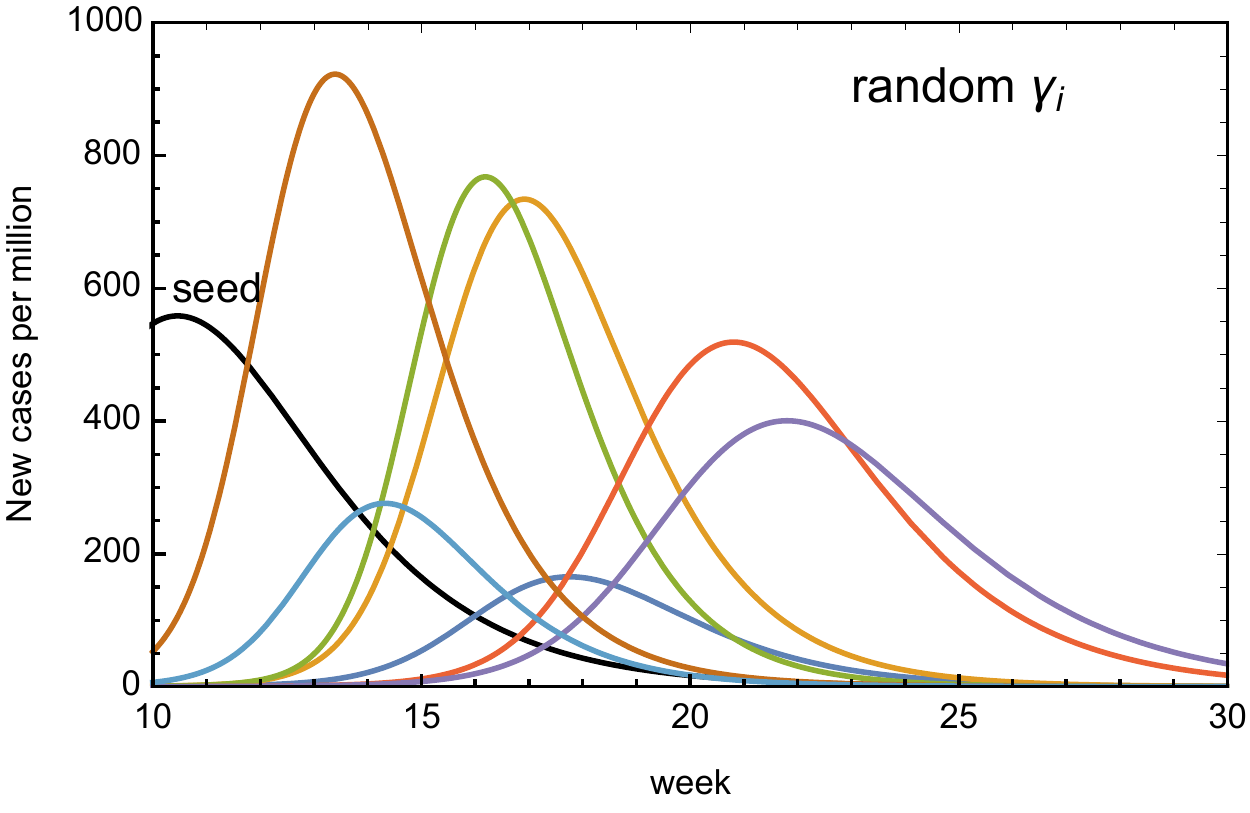}
\includegraphics[width=7cm]{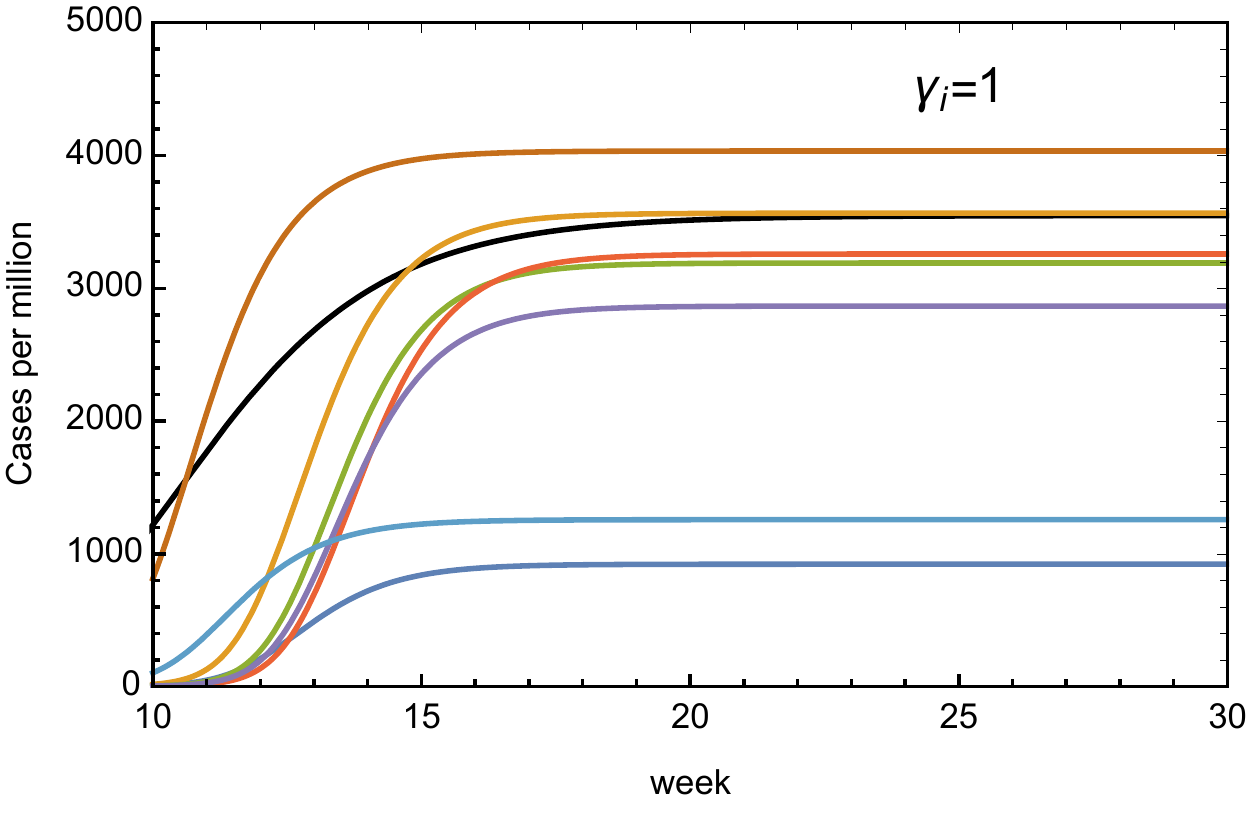}\hspace{1cm}
\includegraphics[width=7cm]{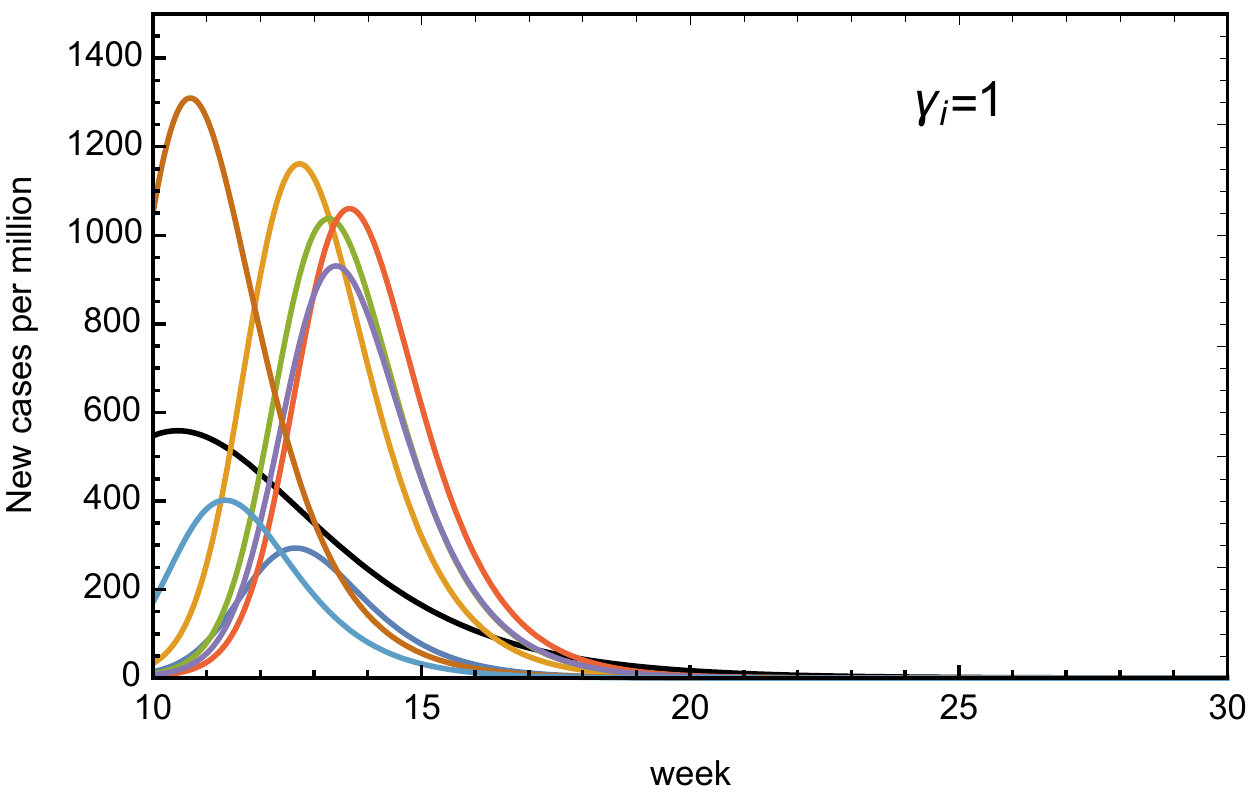}
\end{center}
\caption{Simulation of an epidemic diffusion in a sample of European countries (see text) starting from a ``seed region'', in black. In the top row,
the $\gamma$ coefficients for the European countries are fixed to random values; in the bottom row, they are all fixed to $\gamma_1 = 1$. The
result shows the importance of social distancing measures within each region with respect to the diffusion due to travel.} 
\label{fig:random}
\end{figure*}

\subsection{From Europe to the United States}

To further test our model we consider the system consisting of Europe as region-1 and the United States as region-2. For simplicity, we modelled Europe
on the European Union (with $n_{m1} = 445$) with parameters from the fit of the epidemic diffusion in Italy (c.f., Table~\ref{tab:agamma}). After setting to zero the 
initial cases in the US, we were able to reproduce the diffusion of the epidemic in region-2 (US) for $k=10$, as shown in the right plot of Fig.~\ref{fig:data}.
While it is still  possible that the large value for $k$ may be interpreted as in the above case, it  has the further effect of distorting the epidemic curve for
 region-1, the EU, thus suggesting that it may be hard to explain the diffusion of the COVID-19 epidemic in the US as originating solely from the EU.

At this stage, we cannot exclude that adjusting the epidemic parameters in the EU could improve the agreement. This exercise, nevertheless, proves the effectiveness of our simple eRG model to describe the diffusion of the epidemic among different regions of the world. A more accurate fit may be obtained if more than one region is included in the analysis, which  will be considered in a future work. 

\subsection{Multiple country system: A new simulated epidemic spread in Europe}

We now use the eRG framework to model the impact of a new wave of epidemic spread of the COVID-19 virus (or a related one) in Europe. To do so, we simulate the effect of transmission among countries in a pool of European countries, namely Italy, Spain, France, the United Kingdom, Germany, Denmark and Switzerland. We also include an unspecified ``seed region'', with a population of $n_{m0} = 50$, which has some initial cases, while no case is initially present for the simulated European countries. This is achieved by setting $b_i = \infty$, where $i=1,\dots 7$ spans over the 7 sample countries mentioned above.

We generate randomly the diffusion factors $\gamma_i$ in the range $[0.4, 0.76]$, based on the data of the current COVID-19 epidemic in Europe, and also generate random values of $a_i$ in the range $[7.5, 8.5]$. This also includes the seed region.
Finally, we provide randomly generated numbers of travellers between each of the regions, including the seed one, giving coupling values $k_{ij}$ in the range $[1, 10] \times 10^{-3}$. We then solve the 8 coupled differential equations:
\beq
\frac{d \alpha_i}{d t} = \gamma_i \alpha_i \left( 1-\frac{\alpha_i}{a_i} \right) +  \sum_{j\neq i} \frac{k_{ij}}{n_{mi}} (e^{\alpha_j - \alpha_i}  -1)\,,
\eeq
where $i,j = 0,\dots 7$ and $\alpha_0$ corresponds to the seed region.
The result is shown in the top row of Fig.~\ref{fig:random}, where black indicates the seed region and the coloured curves correspond to the 7 sample European countries.
The top-right plot, where the distribution of new cases is displayed, clearly shows that the peaks in the infected regions occur between 3 to 12 weeks  after the peak in the seed region.
This effect, however, is mainly due to the values of the $\gamma$'s in those regions, and not on the values of the interaction couplings $k_{ij}$.

To prove this, we have run the same simulation again, by fixing $\gamma_i = 1$, $i=1,\dots 7$, while $\gamma_0$ for the seed region is left the same. All other parameters are kept to the same values for the previous case. The analogous results are shown in the bottom row plots of Fig.~\ref{fig:random}. This case roughly correspond to unrestricted diffusion of the virus in the target regions. The result shows that all the peaks are now occurring within 4 weeks after the peak in the seed region.

The results nicely demonstrate that our eRG framework not only is useful, simple and effective to understand the current pandemic, but can also be used to model future ones. 

\section{Conclusions and Discussion}
\label{sec:5}
We extended the epidemic renormalisation group approach to analyse the dynamics of disease transmission and spreading across different regions of the world. We have shown that the eRG framework constitutes an effective way to understand the relative impact of border control versus social distancing measures on the global spread of the epidemic. The  simplicity of the approach, stemming from an effective description of complex phenomena, make it a reliable alternative to the use of expensive high-performance numerical computations.

We calibrated our approach via internationally reported cases.  The approach  elucidates the underlying mechanism that governs the delay in the relative peaks of newly infected cases across different regions of the world.  Among our results, we were able to demonstrate that social distancing measures are more efficient than border control in delaying the epidemic peak. 

In order to connect with widely used time-honoured compartmental models of the SIR-like type, we established the proper map with our eRG framework. We have also shown how to generalise the eRG framework to account for the epidemic interactions across multiple regions of the world. 

We foresee a number of future applications and extensions of our seed work. From a more phenomenological point of view, of immediate impact for society, we plan on embarking on a world-wide monitoring to make global projections that will help governments and industries make containment plans and strategise about reopening society and how to best implement border control. We also wish to improve on understanding the link between the eRG approach and microscopic models of population dynamics and epidemic spread including a number of granular effects that are, by construction, averaged over by effective descriptions such as the eRG approach.

\subsection*{Acknowledgements}
 We thank Michele Della Morte,  Christian M{\o}ller Dahl and Domenico Orlando for comments and helpful discussions.

\newpage

\appendix

\section{Asymptotic behaviour for large $k$} \label{app:largeK}

The coupled system of beta functions in Eqs~\eqref{eq:beta1} and~\eqref{eq:beta2} has an interesting solution in the limit for large $k$.
While this limit should be considered unphysical, it is intriguing from a mathematical point of view. Moreover, it may lead to some
insights on the dynamics of two regions merging into a single one.

In the limit of large $k\to\infty$, the interaction term in the two beta functions dominates.
It thus forces $\alpha_1 = \alpha_2$ asymptotically. To check if there is a fixed point, we can take the sum of the two functions, eliminating the $k$-term, and search for zeros:
\begin{widetext}
\beq
n_{m1} e^{\alpha_1} \beta (\alpha_1) + n_{m2} e^{\alpha_2} \beta (\alpha_2) = e^{\alpha_1} n_{m1} \gamma_1 \alpha_1 \left( 1-\frac{\alpha_1}{a_1} \right) + e^{\alpha_2} n_{m2} \gamma_2 \alpha_2 \left( 1-\frac{\alpha_2}{a_2} \right) = 0\,.
\eeq
\end{widetext}
Imposing the condition $\alpha_1 = \alpha_2 = \alpha$, we find:
\beq
e^\alpha \alpha \left[ n_{m1} \gamma_1  \left( 1-\frac{\alpha}{a_1} \right)  +n_{m2} \gamma_2  \left( 1-\frac{\alpha}{a_2} \right) \right] = 0
\eeq
that is solved by $\alpha = 0$ (UV fixed point) or
\beq
\alpha^\ast = \frac{(n_{m1} \gamma_1 + n_{m2} \gamma_2) a_1 a_2}{n_{m1} \gamma_1 a_2 + n_{m2} \gamma_2 a_1}\,,
\eeq
which defines the new IR fixed point. It is now interesting to ask: for fixed $a_1$ and $a_2$, is the total number of cases in the two regions in the $k\to \infty$ limit larger or smaller than that in the $k\to0$ limit?
In fact, 
\beq
 \mathcal{I}^{\rm tot}_{k\to \infty} &=& (n_{m1} + n_{m2}) e^{\alpha^\ast} = (n_{m1} + n_{m2}) e^{\frac{A^2 - \delta^2}{A + g \delta}} \nonumber \\
&=& (n_{m1} + n_{m2}) e^A e^{- \frac{\delta (\delta + g A)}{A+g \delta}}\,,
\eeq
where we have defined
\beq
\begin{array}{c} A = \frac{a_1+a_2}{2}\,, \;\; \delta = \frac{a_1 - a_2}{2}\,, \\
 g = \frac{n_{m2} \gamma_2 - n_{m1}\gamma_1}{n_{m2}\gamma_2 + n_{m1}\gamma_1}\,.
\end{array}
\eeq
Similarly,
\beq 
\mathcal{I}^{\rm tot}_{k\to0} &=& n_{m1} e^{a_1} + n_{m2} a^{a_2} \nonumber \\
&=& e^A \ (n_{m1} e^\delta + n_{m2} e^{-\delta})\,.
\eeq

\begin{figure}[tb]
\begin{center}
\includegraphics[width=8cm]{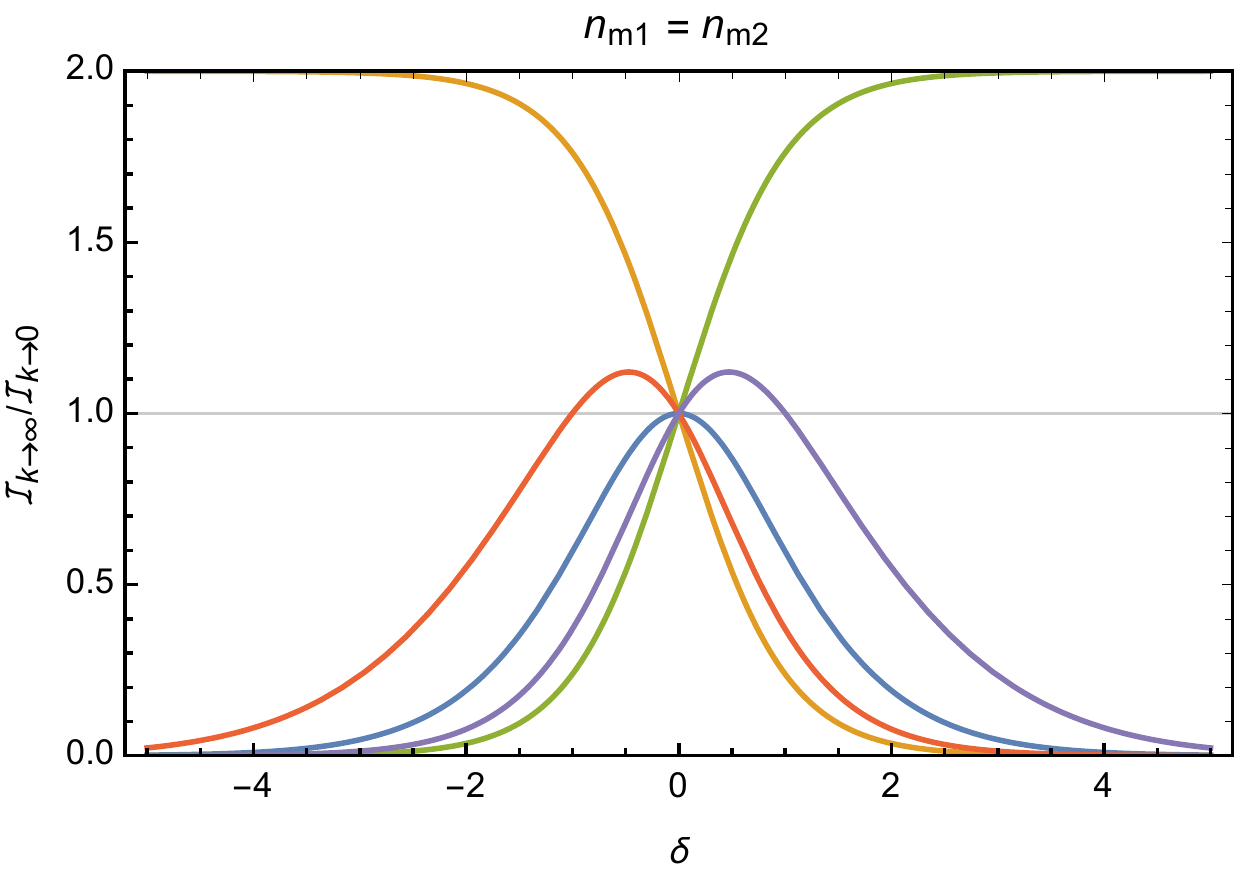}
\includegraphics[width=8cm]{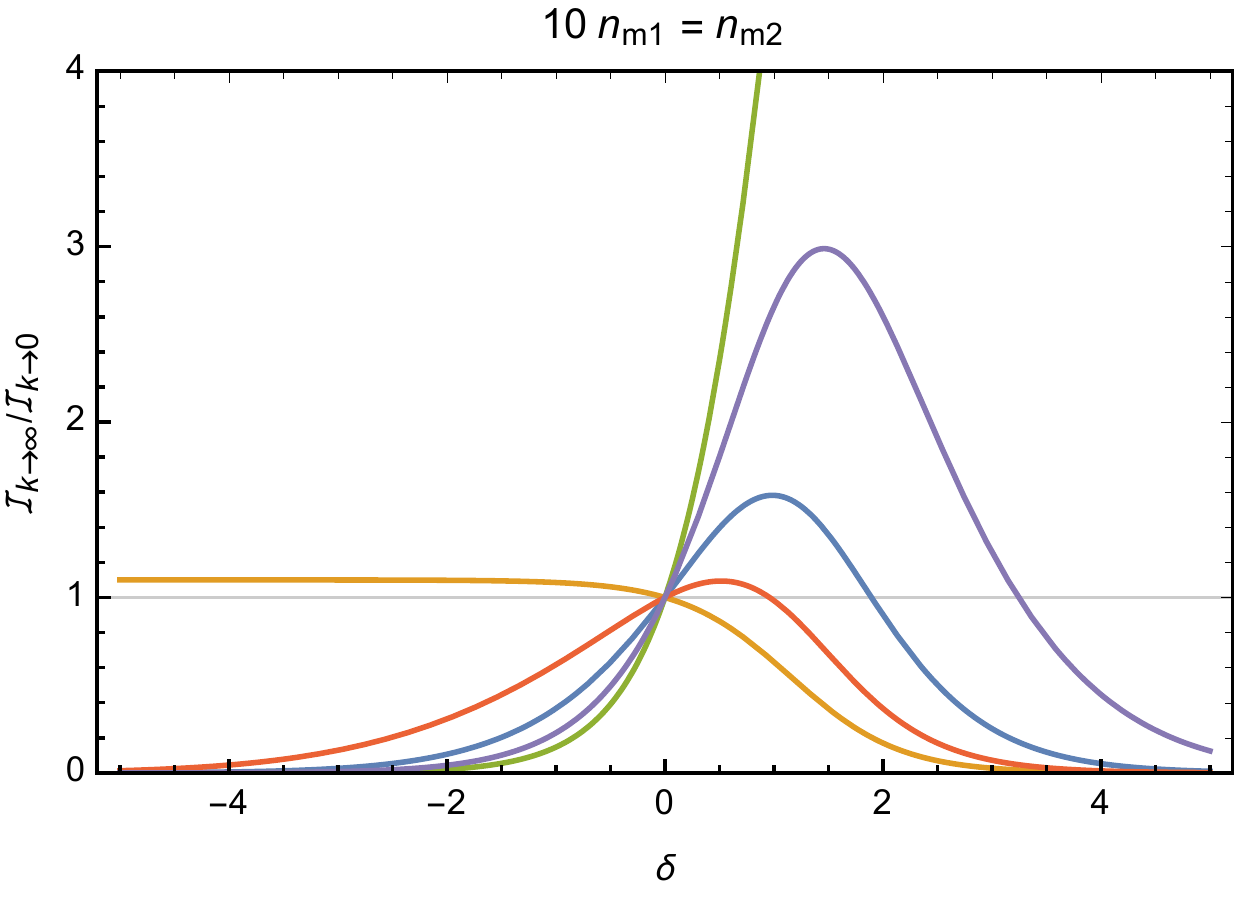}
\end{center}
\caption{Ratio $\mathcal{I}^{\rm tot}_{k\to\infty}/\mathcal{I}^{\rm tot}_{k \to 0}$ as a function of delta for $g=0$ (blue), $g=0.5$ (red), $g=-0.5$ (purple), and the extreme values $g=1$ (orange) and $g=-1$ (green). We fixed $A=12$, but the results have a mild dependence on its value.} \label{fig:LargeK}
\end{figure}

Thus
\beq
\frac{\mathcal{I}^{\rm tot}_{k\to \infty}}{\mathcal{I}^{\rm tot}_{k\to0}} = \frac{n_{m1} + n_{m2}}{n_{m1} e^\delta + n_{m2} e^{-\delta}} e^{- \frac{\delta (\delta + g A)}{A+g \delta}}\,.
\eeq
Interestingly, as the exponential in the numerator may be negative, it is possible to have a reduction in the total number of cases when the exchange of people is large.
In Fig.~\ref{fig:LargeK} we show numerically the above ratio, for $n_{m2} = n_{m1}$ (top plot) and $n_{m2} = 10\, n_{m1}$ (bottom plot), and for various values of $g$
(we recall that $-1 < g < 1$). To make sense of the result, let's consider the case of similar slope in the two countries, i.e. $\gamma_1 = \gamma_2$. In the case $n_{m1} = n_{m2}$, this would give $g = 0$, i.e. the blue curve in the top plot: this case features a reduction of the total cases, as long as $a_1 \neq a_2$.
For $n_{m2} = 10\, n_{m2}$, we have $g \approx 1$, corresponding to the orange curve in the bottom plot: again, there is a significant reduction for $a_1 > a_2$, while the increase if minor if $a_2 > a_1$. Of course, these results are just illustrative, because the values of $a_i$ and $\gamma_i$ should be influenced by the policies concerning the movement of people between infected regions.


\newpage



\begin{thebibliography}{99}



\bibitem{Wilson:1971bg} 
  K.~G.~Wilson,
  ``Renormalization group and critical phenomena. 1. Renormalization group and the Kadanoff scaling picture,''
  Phys.\ Rev.\ B {\bf 4}, 3174 (1971).
  
\bibitem{Wilson:1971dh} 
  K.~G.~Wilson,
  ``Renormalization group and critical phenomena. 2. Phase space cell analysis of critical behavior,''
  Phys.\ Rev.\ B {\bf 4}, 3184 (1971).
  
\bibitem{Sannino:2020epi} 
M.~Della Morte, D.~Orlando and F.~Sannino, ``Renormalization Group Approach to Pandemics: The COVID-19 Case'',  
Frontiers in Physics, Vol 8, 144 (2020), 
\\ \href{https://www.frontiersin.org/article/10.3389/fphy.2020.00144}{Online here}. 
 

%
%
%
%

\bibitem{LI2019566} 
L.~Li, J.~Zhang, C.~Liu, H.T.~Zhang, Y.~Wang and Z.~Wang, 
``Analysis of transmission dynamics for Zika virus on networks'', 
Applied Mathematics and Computation, 347, 566 - 577. 2019. 
 
 \bibitem{ZHAN2018437}
X.X.~Zhan, C.~Liu, G.~Zhou, Z.K.~Zhang, G.Q.~Sun, J.J.H.~Zhu and Z.~Jin, 
``Coupling dynamics of epidemic spreading and information diffusion on complex networks'', 
Applied Mathematics and Computation, 332, 437 - 448, 2018. 

 \bibitem{Perc_2017}
M.~Perc, J.J.~Jordan,  D.G.~Rand, Z.~Wang, S.~Boccaletti and A.~Szolnoki, 
``Statistical physics of human cooperation'', 
{\it Physics Reports} 687, {1--51},   {2017}. 
 
  \bibitem{WANG20151}
 Z.~Wang, M.A.~Andrews, Z.X.~Wu, L.~Wang and C.T.~Bauch, 
 ``Coupled disease--behavior dynamics on complex networks: A review'', 
 Physics of Life Reviews, 15, 1 - 29, 2015. 

 \bibitem{WANG20161}
Z.~Wang, C.T.~Bauch, S.~Bhattacharyya, A.~d'Onofrio, P.~Manfredi, M.~Perc, N.~Perra, M.~Salathe and D.W.~Zhao,  
``Statistical physics of vaccination'', 
Physics Reports, 664, 1 - 113, 2016. 
 
\bibitem{Danby85}
J.M.A.~Danby, 
``Computing applications to differential equations modelling in the physical and social sciences'', 
Reston, Va.: Reston Publishing Company, 1985.

\bibitem{Brauer2019}
F.~Brauer, 
``Early estimates of epidemic final sizes'', 
Journal of Biological Dynamics 13 (sup1):23-30. 2019. 

\bibitem{Miller2012}
J.C.~Miller,  
``A note on the derivation of epidemic final sizes'',
Bulletin of mathematical biology 74 (9):2125-2141. 2012 

\bibitem{Murray}
J.D.~Murray,  
``Mathematical biology'', 3rd ed, Interdisciplinary applied mathematics. New York: Springer. 2002.

\bibitem{Fisman2014}
D.~Fisman, E.~Khoo and A.~Tuite,
``Early Epidemic Dynamics of the West African 2014 Ebola Outbreak: Estimates Derived with a Simple Two-Parameter Model'',
PLOS Currents Outbreaks, 2014 . 


\bibitem{Pell2018}
B.~Pell, K.~Yang, C.~Viboud and G.~Chowell,
``Using phenomenological models for forecasting the 2015 Ebola challenge'',
Epidemics 22:62-70, 2018. 


\bibitem{Kermack:1927}
W.O.~Kermack  and  A.G.~McKendrick,
``A contribution to the mathematical theory of epidemics'',
Proceedings of the Royal Society A. 115 (772): 700â--721. 


  \bibitem{Chinazzi}
  M.~Chinazzi, J.T.Davis, M.Ajello, {\it et al.}, 
  ``The effect of travel restrictions on the spread of the 2019 novel coronavirus (COVID-19) outbreak'', 
  {\it Science} 10.1126/science.aba9757 (2020).
  


  \bibitem{SEIR}
  K.~Prem, Y.~Liu, T.W.~Russell, {\it et al.}, 
  ``The effect of control strategies to reduce social mixing on outcomes of the COVID-19 epidemic in Wuhan, China: a modelling study'', 
 {\it Lancet Public Health}. 2020;S2468-2667(20)30073-6. 
  
  
  \bibitem{scala2020}
 A.~Scala, A.~Fiori, A.~Spelta, {\it et al.}, 
 ``Time, Space and Social Interactions: Exit Mechanisms for the Covid-19 Epidemics'', 
 arXiv:2004.04608 (2020).
  

%
%
  
  

  
  
\end{thebibliography}
\end{document}